# Modélisation et manipulation de données historisées et archivées dans un entrepôt orienté objet

Modelling and querying temporal and archive data
in an object-oriented warehouse

**Franck RAVAT, Olivier TESTE**

IRIT/SIG - 118, route de Narbonne - 31062 Toulouse cedex 04
E-mail : {ravat, teste}@irit.fr

**Résumé**

*Cet article aborde la modélisation et la manipulation des entrepôts objet intégrant des données historisées et archivées. Dans un premier temps, nous proposons un modèle décrivant l'entrepôt comme un référentiel centralisé de données complexes et temporelles. Notre modèle intègre les concepts d'objet entrepôt et d'environnement. Un tel objet est composé d'un état courant, de plusieurs états passés (modélisant les évolutions détaillées) et de plusieurs états archivés (modélisant les évolutions de manière résumée). Le concept d'environnement définit les parties temporelles dans le schéma de l'entrepôt avec une granularité pertinente (attribut, classe, graphe). Dans un second temps, nous définissons une algèbre de manipulation des données de l'entrepôt. Cette algèbre basée sur une extension des algèbres objet propose des opérateurs temporels et des opérateurs de manipulation d'ensembles d'états des objets entrepôt. Une contribution importante réside dans la proposition d'opérateurs spécifiques de restructuration en série temporelle et des opérateurs facilitant les traitements analytiques.*

**Mots clés :** *Entrepôt de Données, BDOO, Données Temporelles et Archivées, Langage de Manipulation.*

**Abstract**

*This paper deals with temporal and archive object-oriented data warehouse modelling and querying. In a first step, we define a data model describing warehouses as central repositories of complex and temporal data extracted from one information source. The model is based on the concepts of warehouse object and environment. A warehouse object is composed of one current state, several past states (modelling value changes) and several archive states (summarising some value changes). An environment defines temporal parts in a warehouse schema according to a relevant granularity (attribute, class or graph). In a second step, we provide a query algebra dedicated to data warehouses. This algebra, which is based on common object algebras, integrates temporal operators and operators for querying object states. An other important contribution concerns dedicated operators allowing users to transform warehouse objects in temporal series as well as operators facilitating analytical treatments.*

# 1 Introduction

De nos jours, les entreprises ont recours à des systèmes décisionnels (OLAP), basés sur l'approche des entrepôts de données [28] pour exploiter d'importants volumes d'information à des fins d'analyse et d'aide à la décision. Un entrepôt [7, 18, 28] permet de stocker les données nécessaires à la prise de décision ; il est alimenté par des extractions de données portant sur des bases opérationnelles, appelées sources de données.

Nos travaux de recherche [20, 21, 22, 23, 26] se placent dans le contexte des systèmes d'aide à la décision basés sur l'approche des entrepôts de données. Nos travaux s'intègrent dans le cadre du projet REANIMATIC[1] et visent plus particulièrement à développer des systèmes décisionnels aptes à supporter efficacement des analyses, afin d'améliorer la qualité des soins et le devenir des patients en réanimation. Notre approche se base sur une dichotomie entre deux espaces de stockage au sein du système décisionnel [3] :

- L'**entrepôt de données** (*data warehouse*) est le lieu de stockage centralisé d'un extrait des bases de production. Cet extrait concerne les données pertinentes pour le support à la décision. Elles sont intégrées et historisées. L'organisation des données est faite selon un modèle qui facilite la gestion efficace des données et leur historisation.
- Le **magasin de données** (*data mart*) est un extrait de l'entrepôt. Les données extraites sont adaptées à une classe de décideurs ou à un usage particulier (recherche de corrélation, logiciel de statistiques,...). L'organisation des données suit un modèle spécifique qui facilite les traitements décisionnels.

L'objet de nos travaux est donc de spécifier des modèles de représentation et des langages de manipulation dédiés aux entrepôts et aux magasins de données complexes et évolutives. Peu de recherches traitent de cette double problématique dans le cadre des systèmes décisionnels : **fournir des modèles de données complexes et évolutives, et spécifier les langages de manipulation associés**.

---

[1] *Le projet **REANIMATIC**, en collaboration avec l'association de médecins OUTCOME-REA, vise à concevoir et à développer un entrepôt de données médicales et évolutives (collectées à partir des bases opérationnelles des services de réanimation) afin d'améliorer la qualité des soins et le devenir des patients dans les services de réanimation des hôpitaux français.*

Nos premiers travaux ont permis de définir l'architecture fonctionnelle d'un système d'aide à la décision [26] distinguant les problématiques de recherche (*cf.* figure 1).

- L'**intégration** se propose de résoudre les problèmes d'hétérogénéité (modèles, formats et sémantiques des données, systèmes,…) des différentes sources de données en intégrant celles-ci dans une source globale. Cette dernière est décrite au moyen du modèle de données orientées objet standard de l'ODMG [6]. Le choix du paradigme objet se justifie car il s'avère parfaitement adapté pour l'intégration de sources hétérogènes [4] couramment utilisées dans le milieu médical [18]. Cette source globale est virtuelle : les données utilisées pour la décision restent stockées dans les sources et sont extraites au moment des mises à jour de l'entrepôt. L'intégration s'appuie sur des techniques de bases de données fédérées [24] et/ou réparties [17].
- La **construction** consiste à extraire les données pertinentes pour la prise de décision, puis à les recopier dans l'entrepôt, tout en conservant, le cas échéant, les changements d'états des données. Par conséquent, l'entrepôt constitue une collection centralisée, de données matérialisées et historisées (conservation des évolutions), disponibles pour les applications décisionnelles. Le modèle de l'entrepôt doit supporter des structures complexes [18] et l'évolution des données au cours du temps [18, 31].
- La **structuration** réorganise l'information dans des magasins de données afin de supporter efficacement les processus d'interrogation et d'analyse, tels que les applications OLAP (*On-Line Analytical Processing*) [8] et la fouille de données (*data mining*) [10] ; les industriels proposent de nombreux outils permettant une telle activité (*Express, Warehouse Builder, Business Object, Impromptu,…*). Pour ce faire, les données importées dans les magasins sont souvent organisées de manière multidimensionnelle [1, 22].

Cette organisation des systèmes décisionnels offre deux visions :

- La **vision informatique**, dédiée à l'entrepôt, permet une gestion et une historisation efficace des données utiles pour les décideurs. L'exploitation directe de l'information n'est réalisable que par des informaticiens.
- La **vision utilisateur**, dédiée aux magasins, permet une représentation des données adaptée aux décideurs. Les analyses et les interrogations sont effectuées de manière indirecte au travers de cette vision multidimensionnelle des données.

Il est important de remarquer que notre entrepôt de données n'est pas organisé de manière multidimensionnelle puisqu'il ne supporte pas directement les processus OLAP. Cette activité est réservée aux magasins de données qui améliorent les performances d'interrogation sans se soucier des redondances d'information ; chaque magasin stocke une partie de l'information disponible dans l'entrepôt afin de répondre à un objectif décisionnel précis ou à un groupe d'utilisateurs ayant les mêmes besoins.

Dans cet article, nous nous focalisons sur l'entrepôt de données. La problématique abordée est double.

- Nous souhaitons proposer un modèle de représentation de l'information intégrant des données complexes et évolutives. L'entrepôt doit stocker uniquement l'information utile pour les décideurs ; l'historisation des données doit être réalisée de manière flexible (sous forme détaillée ou résumée) afin de conserver les seules évolutions utiles.
- Nous souhaitons également proposer un ensemble d'opérateurs de manipulation et d'interrogation des données entreposées. Ces opérateurs doivent permettre d'effectuer des manipulations sur les données courantes et passées.

La section 2 présente les principaux travaux traitant de la modélisation des entrepôts complexes et temporels. La section 3 décrit le modèle de données que nous définissons pour les entrepôts. La section 4 définit l'algèbre de manipulation et d'interrogation de l'entrepôt.

FIG. 1 - Principes d'exploitation du système décisionnel.

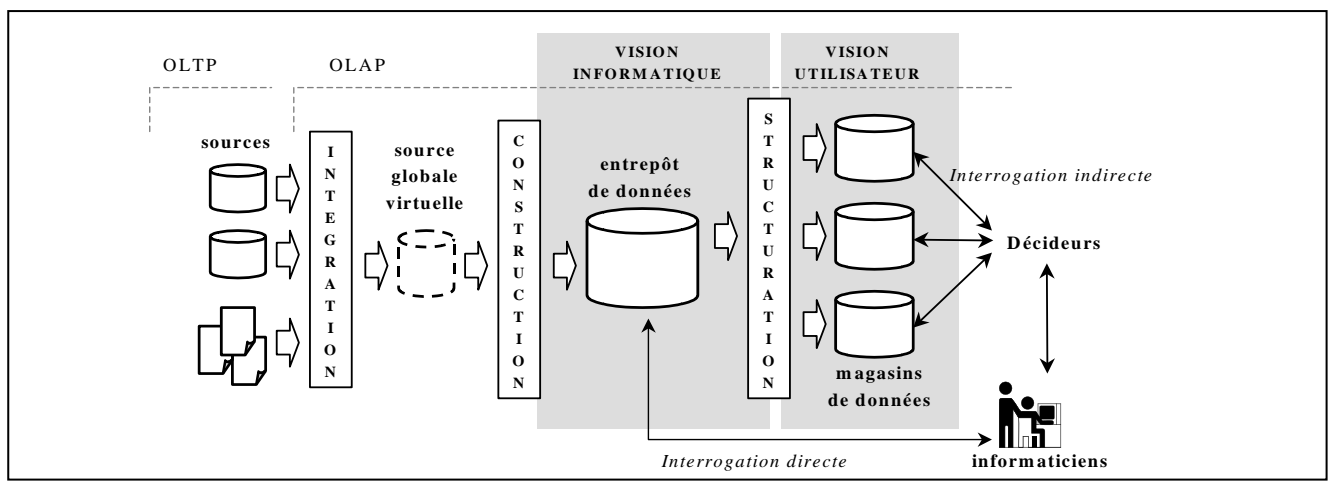

## 2 Limites des approches actuelles

Les travaux relatifs aux entrepôts de données abordent globalement deux problématiques.
- La première traite essentiellement de l'organisation des données. Cette organisation dite multidimensionnelle [1, 13, 15, 19, 22] vise à supporter efficacement les analyses OLAP en offrant une vision des données adaptées et les temps de réponse sont accélérés en calculant de nombreux pré-agrégats.
- La deuxième étudie principalement la sélection et la maintenance incrémentale des vues matérialisées [12, 27, 28, 30, 31, 32] afin de collecter les données utiles aux décideurs, de les stocker dans l'entrepôt et de les maintenir cohérentes avec les données source. Ces travaux se focalisent sur des aspects physiques ou logiques (vues, index,...).

Cependant, ces travaux ne sont pas suffisants et les aspects plus conceptuels restent peu étudiés [11]. Des limites subsistent dans l'élaboration d'un système décisionnel basé sur la dualité entre un entrepôt et des magasins de données.

Les travaux actuels se basent sur des modèles relationnels ou multidimensionnels qui n'intègrent ni des structures complexes ni une sémantique riche. Ceci oblige les concepteurs à un effort d'abstraction important afin de représenter le monde réel dans l'entrepôt. De nouvelles propositions vont dans ce sens ; notamment, [19] définit un modèle orienté objet multidimensionnel pour données complexes et temporelles. Néanmoins, ce modèle n'est pas adapté au niveau de notre entrepôt ; aucun langage de manipulation des données n'est proposé. L'approche multidimensionnelle est adaptée à l'interrogation et l'analyse des magasins de données, mais reste inadéquate pour maintenir et gérer efficacement les données d'un entrepôt sur de longues périodes de temps.

Les propositions actuelles concernant la gestion du temps dans les entrepôts [16, 31] n'offrent pas de mécanismes flexibles pour l'historisation des données ; le plus souvent, les données anciennes sont simplement supprimées de l'entrepôt. Or, aucune proposition ne fournit de mécanismes intermédiaires permettant, par exemple, d'archiver automatiquement les données détaillées à un niveau plus élevé exploitable par les décideurs. [16] fournit un modèle multidimensionnel et un langage temporel T-OLAP de manipulation des données. Ces travaux sont relatifs aux magasins de données (dans notre architecture) et ne paraissent pas adaptés pour un entrepôt dont le modèle de représentation des données ne suit pas une organisation multidimensionnelle. En outre, aucun mécanisme d'archivage n'est proposé.

## 3 Modèle temporel orienté objet

Dans cette section, nous définissons un modèle de données pour les entrepôts, basé sur le paradigme objet. Notre modèle subit l'influence du modèle objet standard de l'ODMG [6] qui est étendu pour prendre en compte les caractéristiques des entrepôts de données. Notamment, notre modèle intègre la dimension temporelle d'une manière flexible en permettant l'archivage des données temporelles.

### 3.1 Objet entrepôt

Chaque information extraite (objet, partie ou groupe d'objets source) est représentée dans l'entrepôt par un **objet entrepôt** qui conserve ses évolutions de valeur au cours du temps (tandis que la source de données ne contient que l'état courant [7], ou bien, ne conserve qu'une partie récente des évolutions, insuffisante pour la prise de décision [31]). Dans un entrepôt, l'administrateur peut décider de conserver :
- l'image de l'information extraite c'est-à-dire l'**état courant**, ainsi que
- les états successifs que prend au cours du temps l'information extraite, c'est-à-dire ses **états passés**,
- uniquement un résumé des états passés successifs, c'est-à-dire l'agrégation de certains états passés, appelée **état archivé**. Les états passés ainsi résumés sont supprimés de l'entrepôt afin de limiter l'accroissement du volume des données.

Un **objet entrepôt** est donc défini par le quadruplet (*oid*, $S_0$, *EP*, *EA*) où *oid* est l'identifiant interne, $S_0$ est l'état courant, $EP = \{S_{p1}, S_{p2},..., S_{pn}\}$ est un ensemble fini contenant les états passés et $EA = \{S_{a1}, S_{a2},..., S_{am}\}$ est un ensemble fini contenant les états archivés.

Un **état** $S_i$ d'un objet entrepôt est défini par le couple $(v_i, h_i)$ où $v_i$ est la valeur de l'objet pour les instants de $h_i$ et $h_i = <[td^1, tf^1[;...;[td^h, tf^h[>$ est le domaine temporel (ensemble ordonné d'intervalles disjoints deux à deux) définissant les instants durant lesquels la valeur de l'état $S_i$ est courante.

La modélisation des domaines temporels s'effectue au travers d'un modèle temporel, linéaire, discret qui définit le temps par le biais d'unités temporelles ; l'espace continu du temps, représenté par une droite de réels, elle-même décomposée en une suite d'intervalles consécutifs disjoints [9]. Chaque partition correspond à une unité temporelle caractérisée par la taille des intervalles décomposant la droite du temps. Notre modèle gère un ensemble d'unités temporelles nommées (*année*, *semestre*, *trimestre*,...) muni d'une relation d'ordre partiel *est-plus-fine* permettant de comparer les unités. Nous définissons plusieurs types temporels de base : l'instant, l'intervalle ainsi que le **domaine temporel**. Ce dernier est un ensemble ordonné d'intervalles disjoints deux à deux et non contigus, noté $h_i = <[td^1, tf^1[; [td^2, tf^2[;...;[td^h, tf^h[>$ où chaque intervalle est non vide ($\forall k \in [1..h], td^k < tf^k$) et possède une même unité temporelle ($\forall k \in [1..h], \forall j \in [1..h], \text{unit}([td^k, tf^k[) = \text{unit}([td^j, tf^j[$ où la fonction *unit*(*Int*) retourne l'unité temporelle de *Int*).

**EXEMPLE :** Nous considérons deux objets entrepôt qui décrivent des patients admis dans un établissement thermal. Nous supposons que différents paramètres sont relevés pour chaque patient (poids, tension…) ; nous avons réduit le nombre de paramètres par rapport à la réalité (dans le projet REANIMATIC, environ 80 paramètres sont définis). Dans notre exemple d'illustration, la périodicité des relevés est fixée au mois (dans les services de réanimation du projet REANIMATIC, les paramètres sont relevés chaque 8h).

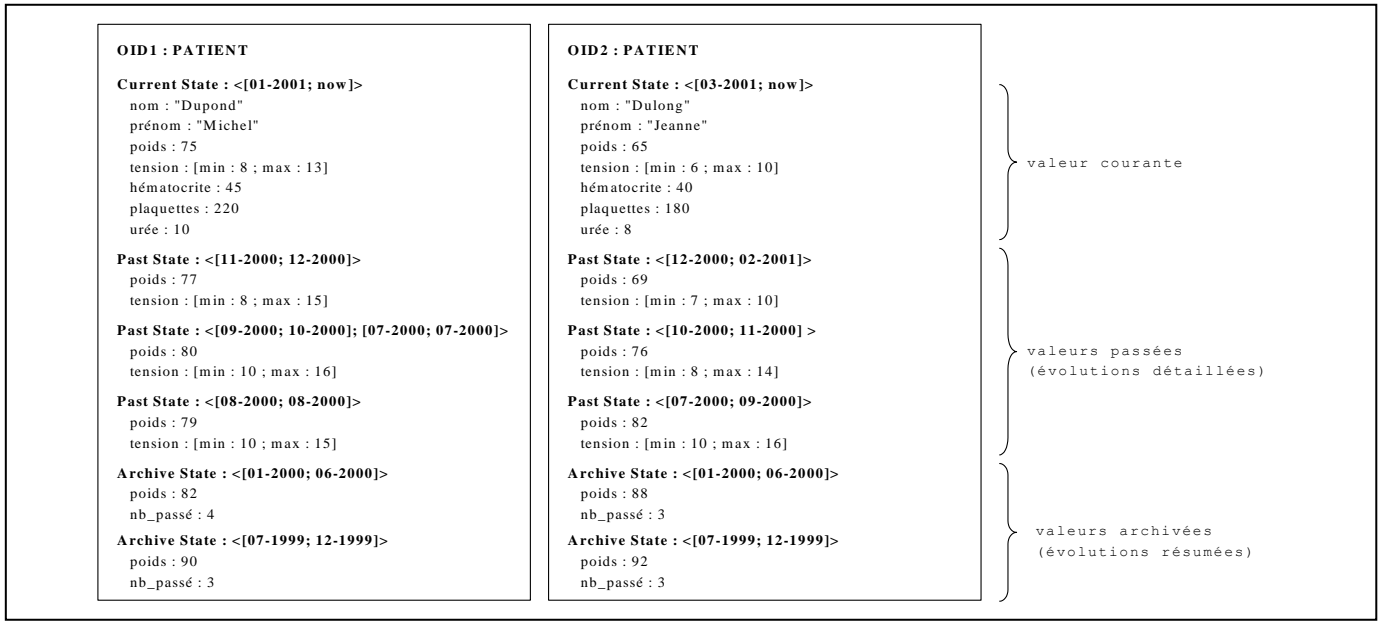

FIG. 2 - Exemple d'objets entrepôt.

## 3.2 Classe entrepôt

### 3.2.1 Définition

Les objets entrepôt qui ont la même structure et le même comportement, sont regroupés dans une classe. Pour prendre en compte les caractéristiques des objets entrepôt, nous définissons le concept de **classe entrepôt** $c$ caractérisé par un n-uplet constitué
- du nom de la classe $Nom^c$,
- d'un type $Type^c$ définissant la structure $Structure^c$ et le comportement $Comportement^c$ des objets entrepôt de $c$ (à chaque classe entrepôt correspond un type),
- d'un ensemble fini de super classes $Super^c$ ($c_i$ est une super classe de $c$, notée $c \preccurlyeq c_i$ si et seulement si, $Type^c \supseteq Type^{c_i}$ et $Extension^c \subseteq Extension^{c_i}$),
- d'une extension $Extension^c = \{o_1, o_2, \ldots, o_x\}$,
- d'une **fonction de construction** $Mapping^c$ qui permet de spécifier le processus d'extraction et de transformation mis en jeu pour créer la structure et le peuplement de la classe $c$ à partir de la source globale,
- d'un **filtre temporel** $Tempo^c$ définissant l'ensemble des propriétés temporelles de $c$ (une propriété est temporelle lorsque ses évolutions sont conservées par des états passés). Le filtre temporel caractérise la structure des états passés des objets de la classe.
- d'un **filtre d'archives** $Archi^c$ définissant l'ensemble des propriétés archivées de $c$ (une propriété est archivée lorsque ses évolutions passées sont résumées dans des états archivés). Le filtre d'archives caractérise la structure des états archivés des objets de la classe.

### 3.2.2 Mécanisme d'extraction

Chaque classe entrepôt est partiellement définie par une fonction de construction $Mapping^c$ appliquée sur la source de données. Cette fonction est une composition de fonctions de base. Nous proposons une taxinomie des fonctions de base supportées par notre modèle d'entrepôt de données répondant aux différents problèmes posés :
- les **fonctions de structuration** (FS) induisent la structure (attributs et relations) des classes entrepôt ;
- les **fonctions de peuplement** (FP) induisent les objets source à partir desquels l'extension des classes entrepôt est calculée ;
- les **fonctions ensemblistes** (FE) correspondent aux opérations ensemblistes classiques de l'algèbre objet de [25] en offrant des mécanismes puissants pour combiner et transformer les classes afin de constituer des classes entrepôt adaptées aux besoins des décideurs ;
- les **fonctions de hiérarchisation** (FH) organisent la hiérarchie d'héritage dans l'entrepôt en créant des super-classes et des sous-classes.

Nous posons $\forall c_i \in C^{ED}$, $Mapping^{c_i} = f^{c_i}_1 \circ f^{c_i}_2 \circ \ldots \circ f^{c_i}_m$ avec $\forall j \in [1,m], f^{c_i}_j \in FS \vee f^{c_i}_j \in FE \vee f^{c_i}_j \in FP \vee f^{c_i}_j \in FH$.

Par souci de simplification, nous ne détaillons pas les différentes fonctions d'extraction ; voir [26] pour une étude détaillée. En outre, nous avons étudié l'extraction du comportement des données dans [20].

**EXEMPLE :** Nous considérons l'exemple précédent. Nous supposons que la source de données à partir de laquelle la classe entrepôt PATIENT a été définie, est composée des classes suivantes :

```
interface Personnes {
    attribute String nom ;
    attribute List<String> prenoms ;
    attribute Boolean sexe ;
    attribute Date naissance ;
    relationship Variables parametres
        inverse Variables::patient ;
    Integer age() ;
}
interface Variables {
    attribute Integer poids ;
    attribute Struct T_tension
            {Integer min, Integer max} tension ;
    attribute Integer hematocrite ;
```

```
    attribute Integer plaquettes ;
    attribute Integer uree ;
    relationship Personnes patient
        inverse Personnes::parametres ;
}
```
A partir de cette source de données, décrite suivant l'ODMG, il est possible de définir la fonction de construction suivante :
```
PROJECT(pp
    JOIN( p Personnes,
          v Variables,
          p.parametres=v ),
    {nom :pp.nom, prénom : pp.prenoms[0],
     poids : pp.poids, tension : pp.tension,
     hématocrite : pp.hematocrite,
     plaquettes : pp.plaquettes, urée : pp.uree})
```
La classe entrepôt produite est définie par :
```
interface PATIENT {
    attribute String nom ;
    attribute String prénom ;
    attribute Integer poids ;
    attribute Struct T_tension
            {Integer min, Integer max} tension ;
    attribute Integer hématocrite ;
    attribute Integer plaquettes ;
    attribute Integer urée ;
}
```

### 3.2.3 Mécanisme d'historisation

Le **filtre temporel** $Tempo^c=\{(p_1, f_1), (p_2, f_2),\ldots, (p_t, f_t)\}$ caractérise les propriétés temporelles d'une classe entrepôt. Il est constitué d'un ensemble de couples $(p_j, f_j)$ où
- $p_j$ est une propriété temporelle et
- $f_j$ est soit un attribut, soit une relation, soit une opération retournant un résultat (fonction).

Les évolutions détaillées des propriétés temporelles sont conservées au travers d'états passés. S'il s'agit d'une opération, les évolutions de son résultat sont conservées à chaque point d'extraction (rafraîchissement de la classe).

**EXEMPLE :** La définition de la classe entrepôt PATIENT est complétée de la manière suivante :
```
interface PATIENT {
    attribute String nom ;
    attribute String prénom ;
    attribute Integer poids ;
    attribute Struct T_tension
            {Integer min, Integer max} tension ;
    attribute Integer hématocrite ;
    attribute Integer plaquettes ;
    attribute Integer urée ;
}
with temporal filter {(poids, poids),
                      (tension, tension)} ;
```
Suivant la définition, les évolutions détaillées du poids et de la tension des patients seront conservées.

### 3.2.4 Mécanisme d'archivage

Le **filtre d'archives** $Archi^c=\{(a_1, f_1), (a_2, f_2),\ldots, (a_s, f_s)\}$ caractérise les propriétés archivées de la classe entrepôt. Il est constitué d'un ensemble de couples $(a_j, f_j)$ où $a_j$ est un attribut et $f_j$ est une fonction d'agrégation. L'ensemble des attributs archivés est un sous-ensemble des attributs temporels. Chaque attribut archivé est associé à une fonction d'agrégation qui définit la manière dont sont résumées les valeurs temporelles.

Les propriétés archivées sont associées à une fonction d'agrégation qui indique comment sont résumées les évolutions détaillées de la propriété temporelle correspondante. Notre modèle supporte plusieurs catégories de fonctions d'agrégation :
- les fonctions d'agrégation forte (*avg, sum, count, max, min*) résument les états passés sélectionnés pour l'archivage dans un seul état archivé ;
- les fonctions d'agrégation modérée (*avg_t, sum_t, count_t, max_t, min_t*) résument les états passés sélectionnés pour l'archivage avec plusieurs états archivés. Les états passés sélectionnés sont regroupés par grain de temps à une unité temporelle supérieure.

FIG. 3 - Comparaison de l'archivage fort et modéré.

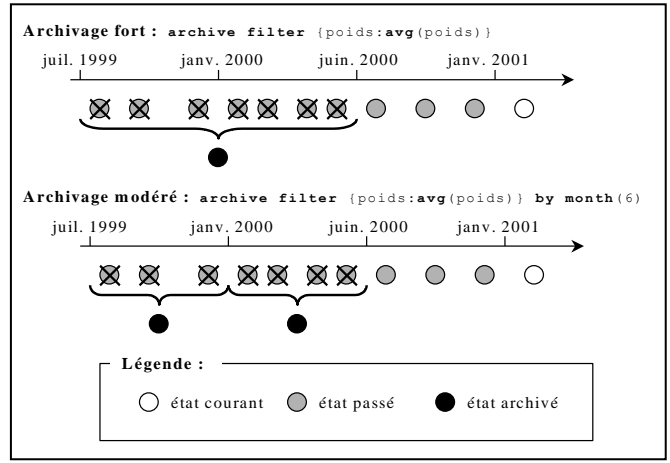

**EXEMPLE :** La définition de la classe entrepôt PATIENT est complétée de la manière suivante :
```
interface PATIENT {
    attribute String nom ;
    attribute String prénom ;
    attribute Integer poids ;
    attribute Struct T_tension
            {Integer min, Integer max} tension ;
    attribute Integer hématocrite ;
    attribute Integer plaquettes ;
    attribute Integer urée ;
}
with temporal filter {(poids, poids),
                      (tension, tension)},
    archive filter {(poids, t_avg(poids))}
    by month(6) ;
```
Suivant la définition, les évolutions détaillées du poids seront archivées (résumées) par périodes de six mois. La mise en place d'un archivage, nécessite un seuil à partir duquel il est déclenché ( se reporter à la section 3.3).

## 3.3 Environnements, schéma de l'entrepôt, configurations

Cette dualité (passé/archive) dans la conservation de l'évolution des données pose une difficulté relative à la définition du comportement temporel des classes. Il est indispensable de fournir des mécanismes permettant :
- de définir des critères pour caractériser les états passés à archiver (seuil de déclenchement de l'archivage),
- de garantir l'intégrité des relations sémantiques (associations, compositions) temporelles. En effet, conserver l'évolution d'une relation exige de conserver les états passés impliqués dans la relation.

Pour cela, nous définissons le concept d'environnement qui regroupe un ensemble de classes entrepôt ayant un comportement temporel homogène (même critère d'archivage, même périodicité de rafraîchissement...). Un **environnement** $Env_i$ est défini par un triplet constitué d'un nom $Nom^{Envi}$, d'un ensemble fini de classes de l'entrepôt $C^{Envi} = \{c^{Envi}_1, c^{Envi}_2, ..., c^{Envi}_{ni}\}$ et d'un ensemble de règles de configuration $Config^{Envi}$ visant à définir le comportement temporel de l'environnement.

Un environnement constitue donc une partie temporellement homogène dans l'entrepôt, ayant ses propres configurations locales $Config^{Envi}$. Nous avons étudié en détail dans [23] les environnements et leurs configurations. Remarquons que ce concept d'environnement aide l'administrateur à définir différentes parties temporelles dans l'entrepôt. Ceci permet de concevoir un entrepôt flexible qui s'adapte aux différentes exigences des décideurs.

Un entrepôt se caractérise par son **schéma** $S^{ED}$ défini par un nom $Nom^{ED}$, l'ensemble fini des classes de l'entrepôt $C^{ED} = \{c_1, c_2, ..., c_n\}$, l'ensemble fini des environnements $Env^{ED} = \{Env_1, Env_2, ..., Env_{ne}\}$ et un ensemble de règles de configuration $Config^{ED}$, visant à définir les différents paramètres de configuration globale de l'entrepôt (période de rafraîchissement,...).

**EXEMPLE :** La classe entrepôt PATIENT possède un filtre temporel et un filtre d'archivage afin de spécifier respectivement les propriétés temporelles et d'archives. Il reste cependant à spécifier le comportement temporel de la classe, et en particulier, il est nécessaire de définir un critère d'archivage qui caractérise les états passés qui doivent être archivés. Nous définissons l'environnement suivant :
**environment** Evolution { PATIENT }
Ensuite, nous spécifions une règle de configuration qui permet d'archiver tous les états passés plus ancien que juin 2000.
```
rule critere_archive on Evolution
when self.refresh()
if select T from P in PATIENT, T in P.PastStates()
   where precedes(T.domT,
                  Date('07-2000', 'mm-aaaa'))
then T.archive() ;
```
Remarquons que la clause if comporte soit une expression booléenne, soit une requête de sélection [29]. Dans notre exemple, la requête permet d'obtenir un ensemble d'états passés sur lequel sera appliqué l'action (si le résultat est vide, l'action n'est pas déclenchée).

### 3.4 Résumé

Nous proposons pour l'entrepôt de données un modèle de représentation orienté objet et intégrant la dimension temporelle de manière flexible au travers un mécanisme d'archivage automatique ; il s'agit d'une extension du modèle standard proposé par l'ODMG. L'originalité de notre modélisation repose sur les concepts d'objets entrepôt, de classe entrepôt et d'environnement.
- L'objet entrepôt est une extension du concept d'objet avec l'intégration de valeurs courantes, mais également celle des valeurs passées sous une forme détaillée (états passés) ou résumée (états archivés).
- Le concept d'objet entrepôt nécessite d'étendre les classes par le concept de classe entrepôt. En plus des caractéristiques standard définissant les classes, une classe entrepôt se caractérise par des filtres temporels et d'archives (pour définir les propriétés temporelles de la classe) et par une fonction de construction (pour définir le processus d'élaboration à partir de la source).
- Le concept d'environnement décrit le regroupement de classes entrepôt ayant un même comportement temporel. L'atout des environnements est qu'ils permettent à l'administrateur de spécifier des parties temporelles homogènes et de taille adéquate aux exigences décisionnelles.

La section suivante se propose de fournir les moyens d'exploiter l'information de l'entrepôt de données.

## 4 Algèbre pour la manipulation des données

Cette section décrit l'algèbre associée au modèle de données préalablement défini. Cette algèbre s'inspire des principales algèbres temporelles objet [5, 25]. Nous reprenons les opérations algébriques des langages pour objets temporels que nous adaptons aux spécificités des objets entrepôt ; ces opérateurs doivent être étendus afin d'intégrer le concept d'état. Nous définissons également de nouveaux opérateurs, spécifiques à notre modélisation, pour manipuler les différentes catégories d'états qui composent les objets entrepôt. Enfin, nous proposons des mécanismes permettant de transformer les données afin d'offrir différentes perspectives sur les évolutions afin de faciliter les traitements analytiques.

### 4.1 Opérateurs classiques

La modélisation spécifique de l'entrepôt réclame une extension des opérateurs proposés dans les langages objets classiques. La principale difficulté se situe au niveau de la prise en compte des mécanismes d'historisation et d'archivage (états courants, passés et archivés).

Ainsi, nous adaptons les principales opérations existantes dans les langages objet aux caractéristiques des objets entrepôts et de leurs états. Le tableau 1 décrit brièvement l'ensemble de ces opérateurs.

On pose T_Object et T_State les types génériques des objets de l'entrepôt et des états ; T_Predicate et T_Attribute représentent respectivement les types de prédicats valides et des attributs définis.

Les opérateurs ensemblistes réalisent l'union, l'intersection ou la différence entre deux ensembles. Ces opérateurs étant basés sur l'égalité, deux types d'union, d'intersection et de différence sont définis : l'un basé sur l'égalité d'identifiant pour les objets entrepôt, l'autre basé sur l'égalité de valeur pour les objets entrepôt ou bien pour les états.

Ces opérateurs peuvent être combinés afin d'exprimer des requêtes complexes à partir d'opérations de base plus simples.

TAB. 1 - Opérateurs classiques.

| OPERATEURS | ENTREE | SORTIE |
|---|---|---|
| VUnion, VIntersect, VDifference | {T_Object}x{T_Object} | {T_Object} |
| | {T_State}x{T_State} | {T_State} |
| IUnion, IIntersect, IDifference | {T_Object}x{T_Object} | {T_Object} |
| Flatten | {{T_Object}} | {T_Object} |
| | {{T_State}} | {T_State} |
| DupElim | {T_Object} | {T_Object} |
| | {T_State} | {T_State} |
| EmptyElim | {{T_Object}} | {{T_Object}} |
| | {{T_State}} | {{T_State}} |
| Select | {T_Object}xT_Predicate | {T_Object} |
| | {T_State}xT_Predicate | {T_State} |
| Project | {T_Object}x{T_Attribute} | {T_Object} |
| | {T_State}x{T_Attribute} | {T_State} |
| Join | {T_Object}x{T_Object}x T_Predicate | {T_Object} |
| | {T_State}x{T_State}x T_Predicate | {T_State} |
| Nest, UnNest | {T_Object}xT_Attribute | {T_Object} |
| | {T_State}xT_Attribute | {T_State} |

## 4.2 Opérateurs d'accès aux états

Nous avons redéfini les opérateurs classiques en fonction de notre représentation de l'information sous forme d'objets entrepôt et d'états, mais il convient de fournir d'autres opérateurs, spécifiques à notre modèle, pour manipuler simplement les différents états des objets entrepôt. Chaque état des objets entrepôt est accessible au travers de trois opérateurs :

- un opérateur Current permettant d'obtenir l'ensemble des états courants tel que Current($\{o_1, o_2,…,o_x\}$)=$\{S^1_0, S^2_0,…,S^x_0\}$ ;
- un opérateur Past permettant d'obtenir un ensemble d'ensembles d'états passés tel que Past($\{o_1, o_2,…,o_x\}$)=$\{Ens^1, Ens^2,…,Ens^x\}$ où $\forall i \in [1..x]$, $Ens^i=\{S^i_{p1}, S^i_{p2},…,S^i_{pi}\}$ est un ensemble d'états passés extraits de l'objet entrepôt $o_i$ ;
- un opérateur Archive permettant d'obtenir un ensemble d'ensembles d'états archivés tel que Archive($\{o_1, o_2,…,o_x\}$)=$\{Ens^1, Ens^2,…,Ens^x\}$ où $\forall i \in [1..x]$, $Ens^i=\{S^i_{a1}, S^i_{a2},…,S^i_{ai}\}$ est un ensemble d'états archivés extraits de l'objet entrepôt $o_i$.

FIG. 4 - Opérateurs d'accès aux états.

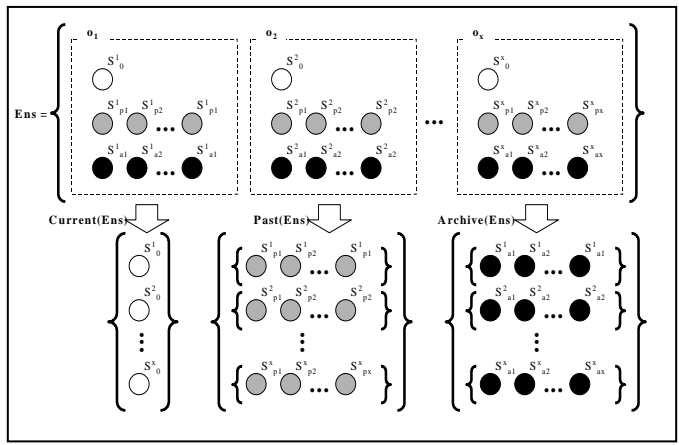

## 4.3 Opérateurs temporels

L'objet de cette section est d'offrir des mécanismes souples et puissants permettant de manipuler simplement la dimension temporelle inhérente aux objets entrepôt.

### 4.3.1 Extension des relations de Allen aux domaines temporels

Afin de faciliter la manipulation des domaines temporels, nous généralisons les treize relations définies par Allen [2] aux domaines temporels associés aux états des objets entrepôt. L'intérêt de cette généralisation est qu'elle offre la puissance nécessaire pour combiner simplement les domaines temporels, c'est-à-dire les valeurs temporelles associées aux états des objets entrepôt.

Soient $X$ et $Y$ deux domaines temporels. On pose
- $X=<[TDeb^X_1, TFin^X_1], [TDeb^X_2, TFin^X_2],…, [TDeb^X_{hx}, TFin^X_{hx}]>$ et
- $Y=<[TDeb^Y_1, Tfin^Y_1], [TDeb^Y_2, Tfin^Y_2],…, [TDeb^Y_{hy}, Tfin^Y_{hy}]>$.

TAB. 2 - Extension des opérateurs de Allen aux domaines temporels.

| RELATION | DEFINITION | RECIPROQUE |
|---|---|---|
| X Precedes Y | $TFin^X_{hx}<TDeb^Y_1$ | Y Follows X |
| X Meets Y | $TFin^X_{hx}=TDeb^Y_1$ | Y IsMeeted X |
| X Overlaps Y | $\exists Int_i \in X, \exists Int_j \in Y$, $TDeb^X_i<TDeb^Y_j \wedge$ $TDeb^Y_j<TFin^X_i \wedge$ $TFin^X_i<TFin^Y_j$ | Y IsOverlaped X |
| X During Y | $\forall Int_i \in X, \exists Int_j \in Y$, $TDeb^X_i>TDeb^Y_j \wedge$ $TFin^X_i<TFin^Y_j$ | Y IsDuringX |
| X Starts Y | $TDeb^X_1=TDeb^Y_1$ | Y IsStarted X |
| X Ends Y | $TFin^X_{hx}=TFin^Y_{hy}$ | Y IsFinished X |
| X Equals Y | $\forall i \in [1,h_1], h_1=h_2$, $TDeb^X_i=TDeb^Y_i \wedge$ $TFin^X_i=TFin^Y_i$ | Y Equals X |

### 4.3.2 Opération de restriction temporelle

Lors de la manipulation directe des données de l'entrepôt, il est souvent nécessaire d'obtenir une information à une date particulière. Autrement dit, un administrateur doit pouvoir obtenir l'état d'un ensemble d'objets entrepôt en fonction d'une valeur temporelle (modélisée soit par un instant précis, soit par un intervalle de temps, soit par un domaine temporel).

Nous définissons l'opérateur State de la manière suivante :
- State(Ens, T, $Op_{Allen}$)=$\{\{S^i_j\} | S^i_j \in o^i \wedge T\ Op_{Allen}\ DomT^i_j\}$ où T désigne le domaine temporel avec lequel sont comparés ceux des états de Ens et $Op_{Allen}$ représente l'opérateur de Allen utilisé pour les comparaisons.

**EXEMPLE :** Nous reconsidérons l'exemple de la classe Patient dont l'extension contient l'objet entrepôt OID1. Un utilisateur qui souhaite se focaliser sur les évolutions du poids et de la tension du patient « Dupond Michel » au cours du second semestre de l'année 2000. Il exprime la requête suivante :

```
State(Select(p Patient,
        p.nom=«Dupond» ∧ p.prénom=«Michel»),
    DomT('07-2000', '01-2001', 'mm-aaaa'),
    during)
```

Le résultat obtenu est le suivant :
```
{[poids=80 ;tension=[min=10 ;max=16] ;
  domT=<[09-2000;10-2000]; [07-2000;07-2000]>] ;
 [poids=79 ;tension=[min=10 ;max=15] ;
  domT=<[08-2000;08-2000]>] ;
 [poids=77 ;tension=[min=8 ;max=15] ;
  domT=<[11-2000;12-2000]>]}
```

### 4.3.3 Opérations de jointure temporelles

La jointure est étendue pour permettre de combiner les ensembles d'états. Cette adaptation de la jointure concerne le caractère temporel des états pour lequel deux possibilités sont offertes :
- le résultat peut être calculé à partir de l'intersection des domaines temporels des états ;
- le résultat peut être calculé à partir de l'union des domaines temporels des états.

Formellement, nous définissons les jointures par intersection et par union des domaines temporels de la manière suivantes :
- IJoin(Ens$^1$, Ens$^2$, P)={(DomT, V$_1$, V$_2$) | ∃S$_1$∈Ens$^1$, ∃S$_2$∈Ens$^2$, DomT=DomT$_1$∩DomT$_2$ ∧ V$_1$∈S$_1$ ∧ V$_2$∈S$_2$ ∧ P(V$_1$, V$_2$)} ;
- UJoin(Ens$^1$, Ens$^2$, P)={(DomT, V$_1$, V$_2$) | ∃S$_1$∈Ens$^1$, ∃S$_2$∈Ens$^2$, V$_1$∈S$_1$ ∧ V$_2$∈S$_2$ ∧ P(V$_1$, V$_2$) ∧ ((DomT⊆DomT$_2$ ∧ V=V$_1$ ∧ (DomT, V)∈Ens$^1$) ∨ (DomT⊆DomT$_1$ ∧ V=V$_2$ ∧ (DomT, V)∈Ens$^2$))}

**EXEMPLE :** Un utilisateur souhaite savoir quand le poids de Dupond Michel était inférieur à celui de Dulong Jeanne (on utilise uniquement les évolutions détaillées). Nous exprimons la requête en utilisant la jointure par union des domaines temporels.
```
UJoin(h1 Flatten(Past(
         Select(p Patient,
         p.nom=«Dupond» ∧ p.prénom=«Michel»))),
      h2 Flatten(Past(
         Select(p Patient,
         p.nom=«Dulong» ∧ p.prénom=«Jeanne»))),
      h1.poids < h2.poids)
```
Le résultat obtenu est constitué de domaines temporels formés par unions des domaines temporels initiaux. La partie structurelle est constituée du poids et d'un couple de valeurs pour la tension conformément à la définition du filtre temporel qui caractérise la structure des états passés (cf. section 3.2.3).
```
{[poids=79 ;tension=[min=10 ;max=15] ;
  domT=<[07-2000;09-2000]>]}
```

### 4.3.4 Opérations de groupements temporels

L'aspect temporel des objets entrepôt permet l'introduction de deux nouveaux opérateurs réalisant des regroupements en fonction de critères temporels :
- un opérateur de groupement se base sur les unités temporelles en permettant de regrouper un ensemble d'états à une unité temporelle supérieure UGroup({S$_1$, S$_2$,…, S$_n$}, U)=Ens'. *Ens'* est un ensemble de n-uplets, dont la valeur est l'ensemble des valeurs structurelles des états *Ens*, regroupés à unité temporelle *U*.
- un opérateur de groupement se base sur la durée en permettant de regrouper des états par durées successives depuis le temps d'origine de l'ensemble des états ; DGroup({S$_1$,…, S$_n$}, D)=Ens'. *Ens'* est un ensemble de n-uplets, dont chaque valeur est l'ensemble des valeurs structurelles des états *Ens*, regroupés par période de temps (exprimée par une durée depuis le plus petit des instants des domaines temporels des états).

## 4.4 Opérateurs de transformation et de traitements analytiques

La prise de décision s'appuie très souvent sur les évolutions organisées chronologiquement [7] ; il s'agit de manipuler ces chronologies en appliquant systématiquement un traitement sur chaque valeur. Or, les opérateurs précédents ne permettent pas d'effectuer ce type de manipulation. Par exemple il est souvent utile d'appliquer une agrégation par accumulation sur des états (qui doivent être organisés chronologiquement) pour obtenir de nouveaux états dont la valeur est calculée en agrégeant les valeurs cumulées des états initiaux. De tels traitements n'ont de sens que sur une série d'états chronologiquement ordonnés.

### 4.4.1 Opération de transformation

Nous définissons donc une **série temporelle** d'états à partir d'un ensemble d'états dont les domaines temporels sont réorganisés sous la forme d'un intervalle. L'ensemble des états est ordonné chronologiquement. Une série temporelle d'états est un ensemble ordonné d'états <S$_1$, S$_2$,…, S$_s$> tel que
- les domaines temporels des états sont des intervalles, ∀i∈[1..s], domT$_i$=<[TDeb$_i$;TFin$_i$]>,
- les états sont ordonnés chronologiquement, ∀i∈[1..s-1], domT$_i$ Precedes domT$_{i+1}$.

Nous proposons un nouvel opérateur de transformation des objets entrepôt en série temporelle. Il s'agit d'un formatage des données sous la forme d'une série temporelle d'états. L'objectif principal de cette transformation est de rendre plus simple la compréhension de l'évolution dans le temps d'un objet entrepôt en permettant l'application de traitements analytiques. L'opérateur MakeSerie transforme un ensemble d'états en une série temporelle d'états. Il se définit de la manière suivante :
- MakeSerie(Ens)=SR où *Ens*={S$_1$, S$_2$, S$_n$} est un ensemble d'états et SR=<S$_{s1}$, S$_{s2}$,…, S$_{ss}$> est une série temporelle d'états.

**EXEMPLE :** Nous considérons l'objet entrepôt OID1. Les états passés de cet objet possèdent des domaines temporels complexes (ils ne peuvent être ordonnés chronologiquement pour simplifier l'analyse). L'utilisateur peut transformer les états passés en une série temporelle d'état ordonnés chronologiquement ; on notera dans la suite SR cette série temporelle.
```
MakeSerie(
  Project(
    pp Flatten(
      Past(
        Select(p Patient,
               p.nom=«Dupond» ∧
               p.prénom=«Michel»)
      )
    ),
    {pp.poids, pp.domT})
)
```

Le résultat obtenu est le suivant :
SR =
```
<[poids=80 ; domT=<[07-2000;07-2000]>] ;
 [poids=79 ; domT=<[08-2000;08-2000]>] ;
 [poids=80 ; domT=<[09-2000;10-2000]>] ;
 [poids=77 ; domT=<[11-2000;12-2000]>]>
```

### 4.4.2  Opérations d'agrégation

Nous adoptons différentes transformations pour les séries temporelles d'états :

- L'opérateur d'agrégation *Agreg* transforme une série temporelle d'états en une valeur ;
  Agreg(SR, Archi)=V où Archi={($att_1$, $f_1$), ($att_2$, $f_2$),…, ($att_t$, $f_t$)} est un filtre d'archivage, c'est-à-dire un ensemble d'attributs (des états de *SR*) associés à une fonction d'agrégation et *V* est la valeur résultat de l'agrégation des valeurs structurelles des états de *Ens*.

**EXEMPLE :** Un utilisateur désire obtenir la moyenne du poids de Dupond Michel. Il ne considère que les états passés (la période considérées concerne donc le second semestre 2000). Il exprime la requête suivante :
**Agreg**(SR, {(poids, avg(poids))})
Le résultat obtenu est le suivant :
[poids=79]

- L'opérateur d'agrégation cumulative consiste à cumuler les résultats d'une agrégation appliquée aux valeurs successives. L'opérateur d'agrégation cumulative *ACum* transforme une série temporelle d'états en une autre série d'états dont les valeurs sont le résultat d'une agrégation cumulée ;
  ACum(SR, Archi)=SR' où Archi={($att_1$, $f_1$), ($att_2$, $f_2$),…, ($att_t$, $f_t$)} est un filtre d'archivage, c'est-à-dire un ensemble d'attributs (des états de *SR*) associés à une fonction d'agrégation et SR'=<$S_{a1}$, $S_{a2}$,…, $S_{aa}$> est une série temporelle d'états dont les valeurs structurelles sont le résultat d'une agrégation cumulée.

**EXEMPLE :** Un utilisateur veut obtenir la moyenne cumulée du poids de Dupond Michel. Il ne considère que les états passés qui correspondent au second semestre 2000.
**ACum**(SR, {(poids, avg(poids))})
Le résultat obtenu est le suivant :
```
<[poids=80   ; domT=<[07-2000;07-2000]>] ;
 [poids=79,5 ; domT=<[07-2000;08-2000]>] ;
 [poids=79,6 ; domT=<[07-2000;09-2000]>] ;
 [poids=79,6 ; domT=<[07-2000;10-2000]>] ;
 [poids=79   ; domT=<[07-2000;11-2000]>] ;
 [poids=79   ; domT=<[07-2000;12-2000]>]>
```

- L'opérateur d'agrégation dynamique permet d'appliquer des agrégations à des ensembles d'états sur une période de temps "*qui se déplace*". Les opérateurs d'agrégation cumulative dynamique *AMove* transforment un ensemble d'états en un autre ensemble d'états dont les valeurs sont le résultat d'une agrégation cumulée par glissements ;
  AMove(SR, Archi, D)=SR' où Archi={($att_1$, $f_1$), ($att_2$, $f_2$),…, ($att_t$, $f_t$)} est un filtre d'archivage, c'est-à-dire un ensemble d'attributs (des états de *SR*) associés à une fonction d'agrégation, *D* est une durée de même unité temporelle que les états de *Ens* et SR=<$S_{a1}$, $S_{a2}$,…, $S_{aa}$> est une série temporel d'états dont les valeurs structurelles sont le résultat d'une agrégation cumulée.

**EXEMPLE :** Un utilisateur veut obtenir la moyenne glissante (avec un décalage de deux mois) du poids de Dupond Michel. Il ne considère que les états passés.
**AMove**(SR,
       {(poids, avg(poids))},
       Duration(2, month))
Le résultat obtenu est le suivant :
```
<[poids=79,5 ;domT=<[07-2000;08-2000]>] ;
 [poids=80   ;domT=<[09-2000;10-2000]>] ;
 [poids=77   ;domT=<[11-2000;12-2000]>]>
```

### 4.4.3  Opérations de changement d'échelle temporelle

Un traitement couramment utilisé lors des analyses est celui du changement de l'échelle d'observation, c'est-à-dire celui du changement de la granularité des domaines temporels des états. Cette transformation peut s'opérer dans deux sens.

- Soit la granularité est augmentée, c'est-à-dire que l'on augmente l'unité temporelle des domaines temporels (cela revient à réduire le détail d'observation des évolutions). On parle de transformation "*scale-up*" ; ScaleUp(SR, $U^+$, Archi)=SR'.
- Soit la granularité est diminuée, c'est-à-dire que l'on diminue l'unité temporelle des domaines temporels (cela revient à augmenter le détail d'observation des évolutions). On parle de transformation "*scale-down*" ; ScaleDown(SR, $U^-$, Archi)=SR'.

**EXEMPLE :** Un utilisateur veut obtenir la moyenne du poids de Dupond Michel avec une échelle temporelle d'observation augmentée du mois au trimestre.
**ScaleUp**(SR,
        'trimestre',
        {(poids, avg(poids))})
Le résultat obtenu est le suivant :
```
<[poids=79,6 ;domT=<[07-2000;09-2000]>] ;
 [poids=78,5 ;domT=<[10-2000;12-2000]>] >
```

## 4.5  Résumé

Le langage de manipulation de l'entrepôt est une extension des algèbres objet prenant en compte les caractéristiques du modèle de représentation de l'entrepôt. L'extension se situe principalement au niveau des opérateurs temporels et des opérateurs de manipulation des ensembles d'états présents dans les objets entrepôt.

Une contribution importante réside dans la proposition d'opérateurs spécifiques de restructuration en série temporelle ainsi que des opérateurs facilitant les traitements analytiques sur les données.

L'intérêt de cette algèbre réside essentiellement dans la base formelle qu'elle offre. Celle-ci peut servir pour le développement futur d'interfaces et de langages graphiques performant permettant la manipulation directe de l'entrepôt dans sa globalité.

TAB. 3 - Synthèse des opérateurs proposés.

| OPERATEURS | DESCRIPTIONS | ORIGINE (*) |
|---|---|---|
| VUnion, VIntersect, VDifference | Opérations ensemblistes basées sur l'égalité de valeur | Ex |
| IUnion, IIntersect, IDifference | Opérations ensemblistes basées sur l'égalité d'identifiant | Ex |
| Flatten | Opération de déstructuration d'un ensemble d'ensembles | Ex |
| DupElim | Opération de suppression des doubles | Ex |
| EmptyElim | Opération de suppression des ensembles vides | Ex |
| Select | Opération de sélection | Ex |
| Project | Opération de projection | Ex |
| Join | Opération de jointure | Ex |
| Nest, UnNest | Opérations de groupement et dégroupement | Ex |
| Current | Opération d'accès aux états courants | Sp |
| Past | Opération d'accès aux états passés | Sp |
| Archive | Opération d'accès aux états archivés | Sp |
| State | Opération d'accès aux valeurs selon une fenêtre temporelle | Sp |
| UJoin | Opération de jointure temporelle par union des domaines temporels | Ex |
| IJoin | Opération de jointure temporelle par intersection des domaines temporels | Ex |
| UGroup | Opération de groupement temporel à une unité temporelle de granularité supérieure | Ex |
| DGroup | Opération de groupement temporel suivant une durée | Ex |
| MakeSerie | Opération de restructuration en série temporelle | Sp |
| Agreg | Opération d'agrégation des séries temporelles | Sp |
| ACum | Opération d'agrégation cumulée des séries temporelles | Sp |
| AMove | Opération d'agrégation glissante des séries temporelles | Sp |
| ScaleUp | Opération de changement d'échelle à une unité temporelle de granularité supérieure | Sp |
| ScaleDown | Opération de changement d'échelle à une unité temporelle de granularité inférieure | Sp |

(*) Les opérateurs proposés sont soit spécifique (Sp) à notre algèbre, soit correspondent à une extension (Ex) d'une opération issue d'algèbres temporelles objet.

# 5 Conclusion

Les travaux de recherche exposés dans cet article se placent dans le contexte des systèmes décisionnels constitués d'entrepôts de données. Notre approche repose sur la dichotomie de deux espaces de stockage au sein du système décisionnel :
- l'entrepôt de données centralise, stocke et historise l'information pertinente pour les décideurs ;
- les magasins de données représentent un extrait de l'entrepôt sous une forme adaptée aux analyses et au support des processus décisionnels.

Ainsi, l'entrepôt de données offre une vision informatique de l'information décisionnelle tandis que les magasins de données proposent une vision utilisateur. Cet article traite plus particulièrement de la vision informatique en proposant un modèle de représentation et une algèbre de manipulation des données de l'entrepôt.

Dans un premier temps, nous avons défini un modèle d'entrepôt de données complexes et temporelles, basé sur le paradigme objet et reposant principalement sur trois concepts.
- Le concept d'objet entrepôt modélise l'état courant d'une information extraite, ainsi que des états passés (représentant les évolutions de l'objet sous une forme détaillée) et des états archivés (correspondant aux évolutions de l'objet décrites sous une forme résumée). L'intérêt de cette modélisation est de conserver les données de l'entrepôt ainsi que leurs évolutions à un niveau de détail pertinent, limitant le stockage des évolutions.
- Le concept de classe entrepôt intègre les caractéristiques de notre approche par une fonction de construction, un filtre temporel et un filtre d'archives.
- Le concept d'environnement permet de définir simplement les parties temporelles homogènes (même période de rafraîchissement, même critère d'archivage,…).

Dans un second temps, nous avons défini une algèbre de manipulation des données de l'entrepôt par extension des principales opérations proposées dans les langages de base de données objet. L'extension se situe essentiellement au niveau des opérateurs temporels et des opérateurs de manipulation des ensembles d'états présents dans les objets entrepôt. En outre, une contribution importante réside dans la proposition d'opérateurs spécifiques de restructuration en série temporelle ainsi que des opérateurs facilitant les traitements analytiques sur les données. L'intérêt majeur de notre algèbre est qu'elle facilite l'interrogation en offrant un cadre générique pour l'interrogation et la manipulation des données de l'entrepôt.

Notre étude fait l'objet d'un développement au travers du prototype GEDOOH[2] [26], acronyme de Générateur d'Entrepôts de Données Orientées Objet et Historisées. Il comporte une interface (visualisant graphiquement la source globale et l'entrepôt de données) et un module générateur (permettant de créer automatiquement des entrepôts ainsi que les processus d'alimentation et de rafraîchissement). GEDOOH est opérationnel : il comprend 8000 lignes de code Java (jdk1.2).

Les perspectives que nous envisageons de conduire sont les suivantes :
- La première problématique que nous allons étudier concerne l'adaptation dynamique des systèmes décisionnels. Actuellement, les techniques utilisées se contentent de proposer une collection de schémas en étoile adaptés aux différents besoins décisionnels. Nous pensons que des profils utilisateurs pourraient être mis à profit pour améliorer l'adaptabilité des systèmes décisionnels en fonction des catégories de décideurs.
- Une seconde problématique concerne la méthodologie de conception des systèmes décisionnels. A l'heure actuelle, aucune méthode de conception n'est disponible pour aider les administrateurs dans leur démarche de conception de ces systèmes. Nous pensons que cette méthode doit s'appuyer sur une étude préalable des méta-données dans les systèmes décisionnels. Ces dernières ont fait l'objet de travaux dans le cadre du projet européen DWQ. Cependant, ces recherches se sont focalisées sur la qualité des systèmes élaborés, sans tenir compte des aspects importants de la démarche et des contraintes de conception.

---

[2] *http://www.irit.fr/SSI/ACTIVITES/EQ_SIG/gedooh.html*

# 6 Références